\documentclass[twocolumn]{aastex631}

\shortauthors{Yoshii et al.}

\usepackage{amsmath,booktabs}	
\begin{document}

\title{Potential signature of Population III pair-instability supernova ejecta \\
in the BLR gas of the most distant quasar at $z=7.54$\footnote{At the
time of writing this paper, the highest redshift record has been updated
to $z = 7.642$, ULAS J0313-1806, as reported by \cite{2021ApJ...907L...1W}.}}

\correspondingauthor{Yuzuru Yoshii}
\email{yoshii@ioa.s.u-tokyo.ac.jp}

\author[0000-0003-3600-5471]{Yuzuru Yoshii}
\affiliation{Institute of Astronomy, School of Science, The University
of Tokyo, 2-21-1 Osawa, Mitaka, Tokyo 181-0015, Japan}
\affiliation{Steward Observatory, University of Arizona, 933 North
Cherry Avenue, Room N204, Tucson, AZ 85721-0065, USA}

\author[0000-0001-6401-723X]{Hiroaki Sameshima}
\affiliation{Institute of Astronomy, School of Science, The University
of Tokyo, 2-21-1 Osawa, Mitaka, Tokyo 181-0015, Japan}

\author[0000-0002-9397-3658]{Takuji Tsujimoto}
\affiliation{National Astronomical Observatory of Japan, Mitaka-shi, Tokyo
181-8588, Japan}

\author[0000-0002-4060-5931]{Toshikazu Shigeyama}
\affiliation{Research Center for the Early universe, Graduate School of
Science, University of Tokyo, Bunkyo-ku, Tokyo 113-0033, Japan}

\author[0000-0003-4573-6233]{Timothy~C. Beers}
\affiliation{Department of Physics and JINA Center for the Evolution of
the Elements (JINA-CEE), University of Notre Dame, Notre Dame, IN 46556,
USA}

\author[0000-0003-3404-0871]{Bruce A. Peterson}
\affiliation{Mount Stromlo Observatory, Research School of Astronomy and
Astrophysics, Australian National University, Weston Creek P.O., ACT
2611, Australia}





\begin{abstract}
The search for Population III  (Pop III) stars has fascinated and 
eluded astrophysicists for decades. One promising place for capturing 
evidence of their presence must be high-redshift objects; signatures 
should be recorded in their characteristic chemical abundances. 
We deduce the Fe and Mg abundances of the broad-line region (BLR) 
from the intensities of ultraviolet \ion{Mg}{2} and \ion{Fe}{2} 
emission lines in the near-infrared spectrum of UKIDSS Large Area 
Survey (ULAS) J1342+0928 at $z = 7.54$, by advancing our novel 
flux-to-abundance conversion method developed for $z\sim 1$ quasars. 
We find that the BLR of this quasar is extremely enriched, 
by a factor of 20 relative to the solar Fe abundance, together with 
a very low Mg/Fe abundance ratio:~$[\mathrm{Fe/H}]=+1.36\pm0.19$ and 
$[\mathrm{Mg/Fe}]=-1.11\pm0.12$, only 700 million years after the 
Big Bang. We conclude that such an unusual abundance feature cannot 
be explained by the standard view of chemical evolution that considers 
only the contributions from canonical supernovae. 
While there remains uncertainty in the high-mass end 
of the Pop III IMF, here we propose that the larger amount of iron 
in ULAS J1342+0928 was supplied by a pair-instability supernova (PISN) 
caused by the explosion of a massive Pop III star in the high-mass end of the possible range of 
$150\textendash 300~M_\sun$.
Chemical-evolution models based on initial PISN enrichment 
well explain the trend in [Fe/Mg]-$z$ all the way from $z < 3$ to 
$z = 7.54$. We predict that stars with very low [Mg/Fe] at all 
metallicities are hidden in the Galaxy, and they will be efficiently 
discovered by ongoing new-generation photometric surveys.
\end{abstract}




\section{Introduction} \label{sec:intro}

According to the Big Bang cosmology, nucleosynthesis does not produce
heavy elements because of the rapid decrease in density and temperature
as the universe expands (\citealt{1967ApJ...148....3W}).  This has led to
an immediate interpretation that the heavy elements observed in various
objects in the universe are synthesized in the interior of massive stars
and ejected by supernovae (SNe).  Therefore, the first generation of
stellar objects called Population III (hereafter, Pop~III) should be
massive stars born from the gas of pristine composition consisting
almost exclusively of hydrogen and helium.

If the initial mass function (IMF) of the hypothetical Pop~III stars
extended to masses as low as $\sim 1~M_\sun$
(\citealt{1977ApJ...211..638S}; \citealt{1977PASJ...29..207S};
\citealt{1980PASJ...32..229Y}; \citealt{1986ApJ...301..587Y};
\citealt{1999ApJ...515..239N,2001ApJ...548...19N};
\citealt{2003ApJ...599..746O}; \citealt{2014ApJ...792...32S};
\citealt{2016ApJ...826....9I}; \citealt{2020ApJ...901...16D}), their
lifetimes would be as long as the age of the Galaxy, and they would
survive to be observed at the present day.  Contrary to expectation,
despite the great observational efforts made during the past four
decades\footnote{Now including spectroscopic samples for hundreds of
thousands, e.g., SDSS/SEGUE; see \cite{2009AJ....137.4377Y} and
\cite{2021R}, and APOGEE; see \cite{2017AJ....154...94M}, to many
millions of stars, e.g., LAMOST; see \cite{2015RAA....15.1095L}.}, no
single star without detectable metals has been found anywhere in the
Galaxy.  Such a null result has been discussed from the perspective as
to whether the low-mass Pop~III stars were misclassified, due to surface
pollution after their birth, as moderately metal-poor stars
(\citealt{1981A&A....97..280Y}; \citealt{1983MmSAI..54..321I};
\citealt{2014ApJ...784..153H}; \citealt{2019MNRAS.486.5917K}), or if
they did not actually form, as suggested by some theoretical arguments
(e.g., \citealt{2004ARA&A..42...79B}; \citealt{2013RvMP...85..809K}). In
any case, the existence of low-mass Pop~III stars will continue to
remain a hypothesis unless a truly zero-metal star is discovered in the
future.

On the other hand, 
while there remains significant uncertainty in our understanding of the high mass end of the Pop III IMF,
massive Pop~III stars with $\gtrsim$10~$M_\sun$ that
cause explosive nucleosynthesis are short-lived and should have already
become SNe long before the Galaxy formed.  However, the
heavy-element abundance pattern of Pop~III SN ejecta should be recorded
in the abundance pattern of second-generation stars born from the
surrounding gas mixed with Pop~III SN ejecta.  Based on this idea,
together with explosive nucleosynthesis calculations of Pop~III SN
progenitors, such nucleosynthesis signatures of massive Pop~III stars
have been searched for in observations of heavy-element abundance
patterns of extremely metal-poor Pop~II stars that are thought to be
second-generation stars nearly as old as Pop~III stars (e.g.,
\citealt{2005Natur.434..871F}; \citealt{2010ApJ...709...11J};
\citealt{2018ApJ...857...46I}).  However, motivated by
the idea that first stars form in small clusters, it has been suggested
that second-generation stars may have formed from gas that was enriched
by multiple Pop~III SNe with different progenitor masses
(\citealt{2018MNRAS.478.1795H,2019MNRAS.482.1204H}).  In this case, the
inverse problem of finding the Pop~III SN progenitor from the abundance
pattern of a second-generation star does not necessarily guarantee the
uniqueness of the solution, and the result is less constrained. In
addition, there remains a problem of whether truly second-generation
stars can be separated solely by their metallicity or age
(\citealt{2018MNRAS.478.1795H}; \citealt{2021MNRAS.506.5410I}).
Investigations along these lines are still ongoing, but the results
obtained so far do not definitively constrain the properties of massive
Pop~III stars in the Galaxy.

It has long been known that if the hydrogen in intergalactic space were
neutral, then no light emitted by quasars at wavelengths shorter than
the wavelength of hydrogen Lyman-alpha (Ly$\alpha$) could be detected
(\citealt{1965ApJ...142.1633G}; \citealt{1995ApJ...444...15Y}).  The
first evidence that the universe is undergoing reionization was found in
observations of the hydrogen Ly$\alpha$ lines in quasar spectra.  The
number of lines decreases with decreasing redshift, indicating that the
reionization is increasing as the universe expands
(\citealt{1978IAUS...79..389P,1983IAUS..104..349P}).  The formation of
quasars and the first Pop~III stars around $z \sim 7\textendash 10$
triggered the transition of the universe from a neutral to fully ionized
state (\citealt{2006AJ....132..117F}; \citealt{2015ApJ...811..140B};
\citealt{2020A&A...641A...6P}).

In recent years, cosmological simulations have inspired attempts to
predict the observability of massive Pop~III stars and their properties
during the epoch of cosmic reionization (\citealt{2016ApJ...823..140X};
\citealt{2018ApJ...854...75S,2019ApJ...871..206S}), and have produced a
growing interest in detecting Pop~III stars through spectroscopic
observations of objects at high redshift that can be traced back in time
to this epoch.  In this context, the near-infrared (NIR) spectrum of the
quasar ULAS J1342+0928 (hereafter, ULAS J1342) at $z=7.54$, which was
recently published by \cite{2020ApJ...898..105O}, has attracted a lot of
attention.  ULAS J1342 is the most distant quasar, and is in the
transition period of cosmic reionization, where quasar
contribution to the reionization is still controversial (e.g.,
\citealt{2015ApJ...813L...8M}; \citealt{2018MNRAS.473.1416M};
\citealt{2022ApJ...929...21}).

This is the first time that a detailed NIR spectrum of the most distant
quasar has been obtained.  The spectrum contains a number of
heavy-element emission lines emitted in the UV-visible wavelength range
of the broad-line region (BLR) in the rest frame.  These lines are
thought to originate from heavy elements supplied to the BLR gas by the
explosive nucleosynthesis of massive stars formed before $z=7.54$.
According to theoretical calculations, the abundance ratio of
$\alpha$-elements such as O, Ne, Mg, Si, and so on, relative to Fe is
sensitive to various types of SN progenitors (e.g.,
\citealt{2002RvMP...74.1015W}; \citealt{2013ARA&A..51..457N}).
Therefore, it is possible to distinguish the SN progenitors from the
[$\alpha$/Fe] measurements with high reliability.  The [$\alpha$/Fe]
abundance ratio for ULAS J1342 at $z=7.54$ is thus expected to provide
new insights into the existence of Pop~III stars.

To investigate the redshift evolution of the
[$\alpha$/Fe] abundance ratio, the flux ratio of \ion{Fe}{2} and
\ion{Mg}{2} emission lines in the quasar spectrum has recently been
measured by many researchers, as summarized in
\cite{2020ApJ...905...51S}.  However, no clear redshift evolution has
been seen in these flux ratio measurements (e.g.,
\citealt{2011ApJ...739...56D}; \citealt{2017ApJ...849...91M};
\citealt{2019ApJ...874...22S}; \citealt{2020ApJ...898..105O};
\citealt{2020ApJ...905...51S}).  On the other hand, statistical studies
of quasar spectra have shown that the \ion{Fe}{2}/\ion{Mg}{2} flux ratio
depends not only on the abundance of heavy elements but also on
non-abundance parameters (\citealt{2011ApJ...736...86D,
2017ApJ...834..203S}).  These results suggest that the conversion of the
\ion{Fe}{2} and \ion{Mg}{2} emission-line fluxes to the [Mg/Fe]
abundance ratio is essential to quantitatively investigate chemical
evolution.  We have recently developed such a conversion method, and
found that the [Mg/Fe] abundance ratio of the BLR is almost constant at
$0.7 < z < 1.6$ (\citealt{2017ApJ...834..203S}).  The systematic errors
of [Mg/Fe] and [Fe/H] arising from the conversion method itself are
estimated from the fluctuation of their averages for each redshift bin,
and are as small as 0.1~dex.  Thus, with our method, we have opened a
window for the study of the chemical evolution of the universe at high
redshift using quasars.

The present paper is the first report to determine the abundance ratio
of [Mg/Fe] from the NIR spectrum of ULAS J1342, and to identify the SN
progenitor based on the flux-to-abundance conversion method.  As a
result, we deduce that the value of [Mg/Fe] for ULAS J1342
originates from a pair-instability supernova (PISN) by a Pop~III star of
several hundred solar masses.  Based on this result, 
we show that the
study of the Pop~III era at $z > 7$ is crucial for understanding the
chemical evolution of the early universe, and we discuss {\color{black}
the potential} of high-redshift quasar surveys for various studies of
Pop~III stars in the future.  Throughout this paper, we assume a
$\Lambda$CDM cosmology, with $\Omega_\Lambda = 0.7$, $\Omega_M = 0.3$,
and $H_0 = 70~\mathrm{km~s^{-1}~Mpc^{-1}}$.


\section{Measurement and Analysis}

\subsection{Abundance Analysis of ULAS J1342+0928} \label{sec:measure}

The measurements of ULAS J1342 are taken directly from
\cite{2020ApJ...898..105O}.  Table~\ref{tab:ULAS_param} shows their
values.  Here we give only a brief summary of how these values were
measured.

The spectrum of ULAS J1342 was taken with the Gemini Near-InfraRed
Spectrograph (GNIRS) at the Gemini North telescope.  The spectrum covers
the observed wavelengths of $0.9\textendash 2.5~\micron$ 
with a spectral
resolution of $R \sim 760$.  The power-law plus Balmer continuum and the
\ion{Fe}{2} template were iteratively fitted to the observed spectrum,
which is a common technique used in this research area (e.g.,
\citealt{2011ApJ...739...56D,2017ApJ...849...91M,2017ApJ...834..203S}).
In \cite{2020ApJ...898..105O}, there are results from two different
\ion{Fe}{2} templates, one by \cite{2001ApJS..134....1V} and one by
\cite{2006ApJ...650...57T}.  In this paper, we adopt the result for the
case where the \ion{Fe}{2} template of \cite{2006ApJ...650...57T} was
used, because \cite{2017ApJ...834..203S,2020ApJ...904..162S}, who
proposed the flux-to-abundance conversion method, also used the same
\ion{Fe}{2} template.  The fitted \ion{Fe}{2} template was integrated
over $\lambda_\mathrm{rest}=2200\textendash 3090$~\AA\ and then divided
by the continuum flux at $\lambda_\mathrm{rest}=3000$~\AA\ to derive the
rest-frame equivalent width (EW).  After subtracting the continuum
components, \ion{Mg}{2} $\lambda2798$ was fitted with a single Gaussian
and the rest-frame EW and {\color{black} the full width of the line at
half maximum} (FWHM) were measured.  The black hole
mass, $M_\mathrm{BH}$, was estimated from the measured monochromatic
luminosity at 3000~\AA\ ($\lambda L_\lambda(3000~\mathrm{\AA}) \equiv
L_{3000}$) and FWHM(\ion{Mg}{2}) based on the single-epoch mass
estimator given by \cite{2009ApJ...699..800V}.  The bolometric
luminosity was measured by applying a bolometric correction of
$L_\mathrm{bol}=5.15 \times L_{3000}$ (\citealt{2006ApJS..166..470R}).
The Eddington ratio was obtained by dividing $L_\mathrm{bol}$ by the
Eddington luminosity, $L_\mathrm{Edd}=1.3 \times 10^{38}
(M_\mathrm{BH}/M_\sun)~\mathrm{erg~s^{-1}}$, which is the classical 
definition for a completely ionized pure hydrogen gas.

\begin{deluxetable}{lr}[t]
\tablecaption{Parameters for ULAS J1342+0928 \label{tab:ULAS_param}}
\tablehead{
 \colhead{Parameter} & \colhead{Value}
}
 \startdata
 \multicolumn{2}{c}{Observation settings} \\ \hline
 Observed wavelength (\micron) & 0.9--2.5 \\
 Spectral resolution ($\Delta\lambda/\lambda$) & 760 \\ \hline
 \multicolumn{2}{c}{Measurement} \\ \hline
 Redshift & 7.54 \\
 $L_{3000}$ ($\mathrm{10^{46}~erg~s^{-1}}$) & $2.47\pm0.03$ \\
 EW(\ion{Mg}{2}) (\AA) & $13.4^{+0.8}_{-0.9}$ \\
 EW(\ion{Fe}{2}) (\AA) & $126^{+6}_{-15}$ \\
 FWHM(\ion{Mg}{2}) ($\mathrm{km~s^{-1}}$) & $2830^{+210}_{-210}$\\ \hline
 \multicolumn{2}{c}{Estimate} \\ \hline
 $M_\mathrm{BH}~(10^8~M_\sun)$ & $7.6^{+3.2}_{-1.9}$ \\
 $L_\mathrm{bol}/L_\mathrm{Edd}$ & $1.5^{+0.5}_{-0.4}$ \\
\enddata
\tablerefs{\cite{2020ApJ...898..105O}, where the adopted
 values of the cosmological parameters are the same as in this paper.}
\end{deluxetable}

\begin{figure}[t]
 \epsscale{1.1}
 \plotone{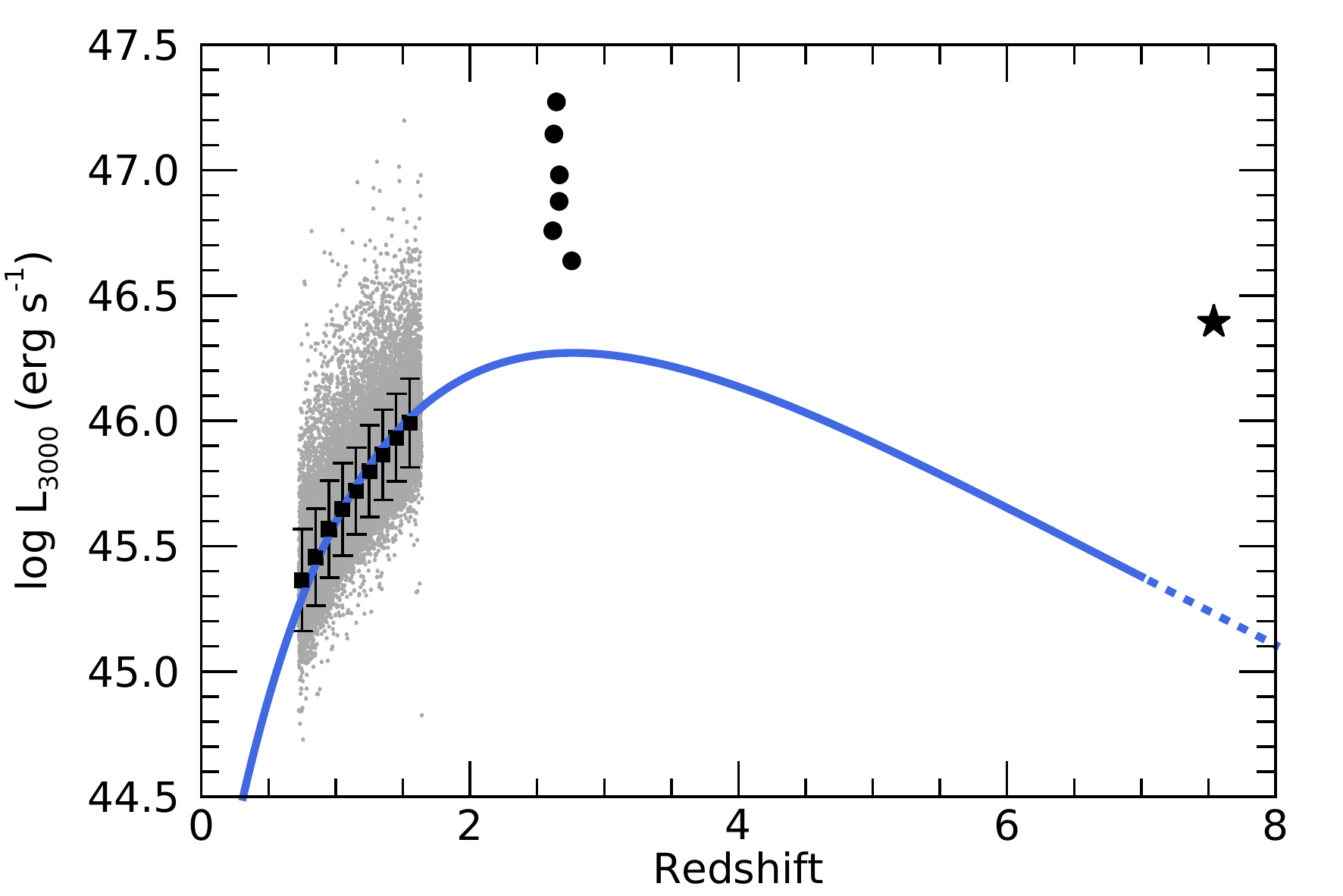}
 \caption{Comparison of the form of $L_*(z)$ in the ``Global fit A''
 model with observed data.  The gray dots indicate the SDSS quasars
 analyzed by \cite{2017ApJ...834..203S}, together with the mean and
 standard deviation for each redshift bin.  The circles indicate the $z
 \sim 2.7$ quasars analyzed by \cite{2020ApJ...904..162S}, and the star
 indicates ULAS J1342.  The blue solid line is $L_*(z)$ of the ``Global
 fit A'' model in \cite{2020MNRAS.495.3252S}, scaled to fit the mean
 $L_{3000}$ of the SDSS quasars, and the blue dashed line is its
 extrapolation to $z > 7$.}  \label{fig:lumievo}
\end{figure}

\begin{figure*}[t]
 \epsscale{1.1}
 \plotone{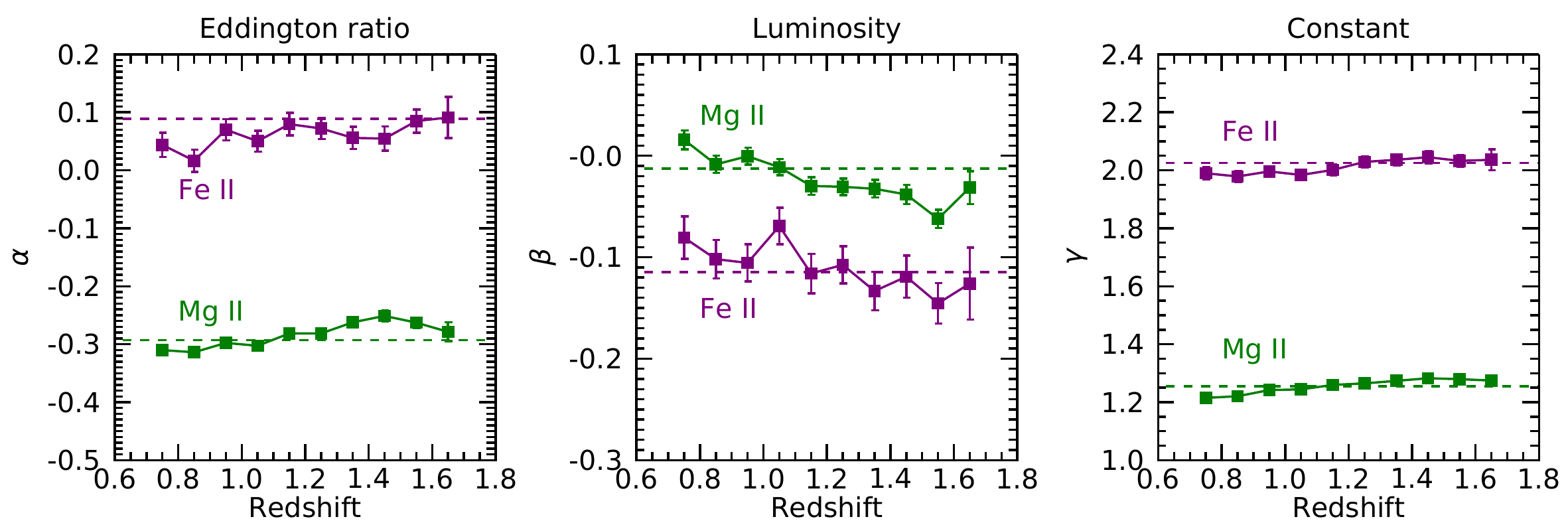}
 \caption{Parameters obtained by fitting the EWs of the SDSS quasars
 measured by \cite{2017ApJ...834..203S} with
 Equation~(\ref{eq:ew_model}) as a function of redshift.  \ion{Mg}{2}
 results are shown in green and and \ion{Fe}{2} results are shown in
 purple.  Dashed lines represent the results of the fitting when all
 SDSS samples are used without binning by redshift.  }
 \label{fig:fitting}
\end{figure*}

In \cite{2020ApJ...904..162S}, we demonstrated that the measured EWs
need to be corrected for the Eddington ratio dependence and Baldwin
effect by the following equation before comparing with the
photoionization model:
\begin{equation}
 \mathrm{EW}^\prime = \mathrm{EW} \left( \frac{L_\mathrm{bol}/L_\mathrm{Edd}}{A}
		          \right)^{-\alpha} \left(
		          \frac{L_{3000}}{B} \right)^{-\beta}.
		          \label{eq:ew_correct}
\end{equation}
Here, $A$ and $B$ are the fiducial values of the Eddington ratio and
monochromatic luminosity at 3000~\AA, respectively.
Because the dependence of the Eddington ratio on redshift is only
slight, the median value of the quasar sample at $0.7 \lesssim z
\lesssim 1.6$ retrieved from the Sloan Digital Sky Survey (SDSS) Data
Release 7, $A=10^{-0.55}$, was used as the fiducial value independent of
redshift.  On the other hand, since the quasar luminosity evolves
significantly with redshift, we adopted the characteristic luminosity
($L_*$) of the quasar luminosity function (QLF) obtained from the
SDSS-III Data Release 9 (\citealt{2013ApJ...773...14R}) as the fiducial
value to estimate [Mg/Fe] and [Fe/H] of quasars at $z \le 2.7$.
However, because the QLF of \cite{2013ApJ...773...14R} covers only up to
$z = 3.5$, it would be inappropriate to extrapolate and apply it to ULAS
J1342.  Recently, \cite{2020MNRAS.495.3252S} compiled observed quasar
data at various wavelengths {\color{black} which were obtained} over the
past decades {\color{black} and are} available up to $z=7$.  The
redshift evolution of $L_*$ in their ``Global fit A'' model is written
as:
\begin{equation}
 \log L_*(z) = \frac{2c_0}{\left( \frac{1+z}{1+z_\mathrm{ref}} \right)^{c_1}
  + \left( \frac{1+z}{1+z_\mathrm{ref}} \right)^{c_2}}, \label{eq:L_S20}
\end{equation}
where $(c_0, c_1, c_2)=(13.0088, -0.5759, 0.4554)$, and
$z_\mathrm{ref}=2$.  Figure~\ref{fig:lumievo} compares the form of this
redshift evolution of $L_*$ with $L_{3000}$ of ULAS J1342 and our
previously analyzed low-redshift quasars
(\citealt{2017ApJ...834..203S,2020ApJ...904..162S}).  In the figure, we
have kept the form of the ``Global fit A'' model, but scaled it to match
$L_{3000}$ of the SDSS sample at $0.7 \lesssim z \lesssim 1.6$.  In the
current situation, where the luminosity evolution of quasars at $z > 7$
is unclear, the most natural approach is to extrapolate and apply the
``Global fit A'' model to ULAS J1342.  Thus, the fiducial luminosity $B$
in Equation~(\ref{eq:ew_correct}) can be written as:
\begin{equation}
 \log B = \frac{2c_0}{\left( \frac{1+z}{1+z_\mathrm{ref}} \right)^{c_1}
  + \left( \frac{1+z}{1+z_\mathrm{ref}} \right)^{c_2}} + c_3, \label{eq:L_S20}
\end{equation}
where $(c_0, c_1, c_2, c_3)=(13.0088, -0.5759, 0.4554, 33.17)$ and $c_3$
corresponds to the scaling factor.  Note that at $z < 3$, the redshift
evolution of $L_*$ is almost the same in both \cite{2013ApJ...773...14R}
and \cite{2020MNRAS.495.3252S}.  Thus, adopting the latter does not
affect the conclusion in \cite{2020ApJ...904..162S}.

The exponents $\alpha$ and $\beta$ in Equation~(\ref{eq:ew_correct}) are
obtained by fitting the SDSS data at $0.7 \lesssim z \lesssim 1.6$ as in
\cite{2020ApJ...904..162S} with the following equation:
\begin{equation}
 \log{\mathrm{EW}} = \alpha \left(
		     \frac{L_\mathrm{bol}/L_\mathrm{Edd}}{A}
		    \right) + \beta \left( \frac{L_{3000}}{B}
			        \right) +
		    \gamma. \label{eq:ew_model}
\end{equation}
The fitting results are shown in Figure~\ref{fig:fitting}.  Both
exponents $\alpha$ and $\beta$ exhibit no significant evolution with
respect to redshift.\footnote{In \cite{2020ApJ...904..162S}, we obtained
fitting results that did not rule out the possibility of redshift
evolution of $\beta$ at $z \gtrsim 1.5$.  However, after rechecking, we
found that this was due to the fact that the least-squares fitting was
heavily influenced by outliers.  In this paper, we performed the fitting
after excluding the obvious outliers.}  For the redshift evolution of
$\beta$ (i.e., the Baldwin effect), it has been reported that there is
no evolution up to $z \sim 5$ in \ion{C}{4} $\lambda1549$
(\citealt{2008MNRAS.389.1703X} (SDSS Data Release 5);
\citealt{2012MNRAS.427.2881B} (SDSS Data Release 7)).  Under these
circumstances, it is natural to assume that $\alpha$ and $\beta$ do not
undergo redshift evolution.  For the SDSS quasars in the range $0.7 < z
< 1.6$, we obtain values of $(\alpha,\beta,\gamma)=(-0.291,-0.028,1.26)$
for \ion{Mg}{2} and $(\alpha,\beta,\gamma)=(+0.086,-0.127,2.04)$ for
\ion{Fe}{2}, which are shown as dashed lines in
Figure~\ref{fig:fitting}.  In the present study, these values are
applied to all samples including ULAS J1342.

By applying the corrected EWs obtained from
Equation~(\ref{eq:ew_correct}) to the abundance diagnostic diagram (see
Figure 9 in \citealt{2020ApJ...904..162S}), we obtained
$([\mathrm{Mg/Fe}], [\mathrm{Fe/H}])=(-1.11\pm0.12, +1.36\pm0.19)$ for
ULAS J1342.  The errors were estimated through Monte Carlo simulations
by randomizing the observables of \cite{2020ApJ...898..105O} based on
their measurement errors.  In addition to ULAS J1342, similarly measured
abundances for the low-redshift quasars analyzed by
\cite{2017ApJ...834..203S,2020ApJ...904..162S} are listed in
Table~\ref{tab:abundance}.

\subsection{Comparison with Previous Studies}

\begin{deluxetable}{lccc}[t]
\tablecaption{Measured abundances \label{tab:abundance}}
\tablehead{
 \colhead{Sample} & \colhead{Redshift} & \colhead{[Mg/Fe]} &
 \colhead{[Fe/H]}
}
 \startdata
 SDSS       & 0.60--0.75 & $-0.30\pm0.30$ & $+0.39\pm0.63$ \\
 SDSS       & 0.75--0.90 & $-0.29\pm0.33$ & $+0.37\pm0.65$ \\
 SDSS       & 0.90--1.05 & $-0.22\pm0.31$ & $+0.31\pm0.59$ \\
 SDSS       & 1.05--1.20 & $-0.18\pm0.24$ & $+0.28\pm0.54$ \\
 SDSS       & 1.20--1.35 & $-0.31\pm0.21$ & $+0.45\pm0.52$ \\
 SDSS       & 1.35--1.50 & $-0.33\pm0.28$ & $+0.49\pm0.58$ \\
 SDSS       & 1.50--1.75 & $-0.25\pm0.31$ & $+0.40\pm0.59$ \\
 NTT        & 2.7        & $-0.10\pm0.41$ & $+0.45\pm0.42$ \\
 ULAS J1342 & 7.54       & $-1.11\pm0.12$ & $+1.36\pm0.19$ \\
 \enddata
\end{deluxetable}

\begin{figure}[t]
 \epsscale{1.1} \plotone{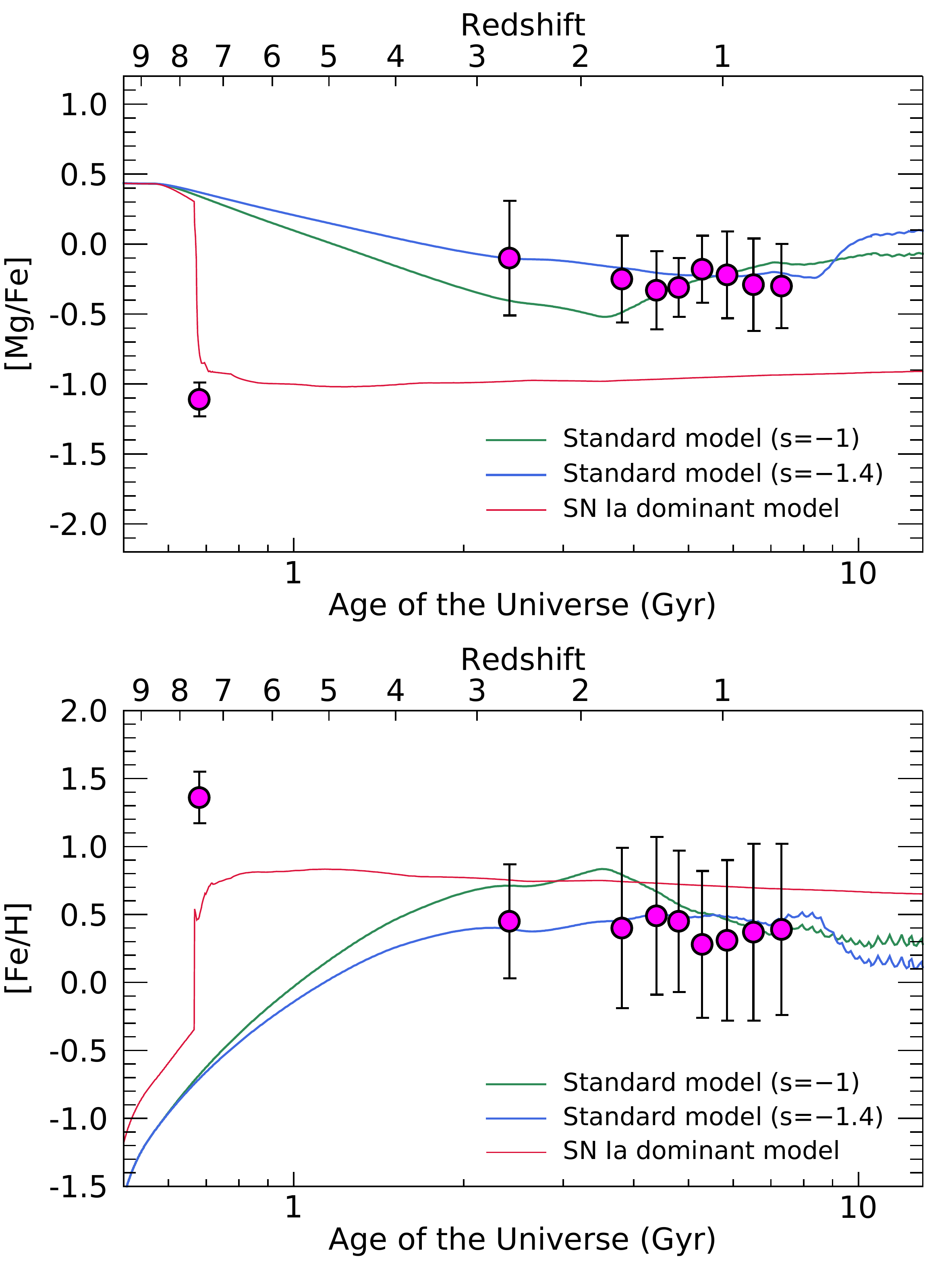}
 \caption{[Mg/Fe] (upper panel) and [Fe/H] (lower panel) as a function
 of the age of the universe, plotted as magenta filled circles for the
 results of ULAS J1342 and $z < 3$ samples from
 \cite{2017ApJ...834..203S,2020ApJ...904..162S}.  The green and blue
 curves are the standard chemical evolution models
 for the core region of 
 an elliptical galaxy, for two cases with
 different delay time distributions of SNe~Ia.  The red curve is the
 variant model in which SN Ia ejecta predominates in the early
 galaxy.}
 \label{fig:stdmodel}
\end{figure}

Figure \ref{fig:stdmodel} shows [Mg/Fe] and [Fe/H] as a function of the
age of the universe, for predictions of chemical evolution models of
elliptical galaxies, which are thought to host quasars, and measurements
of quasars from \cite{2017ApJ...834..203S,2020ApJ...904..162S}.  The
details of the chemical evolution models will be described in the next
subsection (\S\ref{sec:std_model}).  In addition to ULAS J1342 at
$z=7.54$, quasar data at $z<3$ from
\cite{2017ApJ...834..203S,2020ApJ...904..162S} are also plotted for
comparison.  Other high-redshift quasar data such as
\cite{2011ApJ...739...56D,2014ApJ...790..145D} and
\cite{2017ApJ...849...91M} are not plotted in this figure because their
results used the \ion{Fe}{2} template of \cite{2001ApJS..134....1V}
instead of \cite{2006ApJ...650...57T}, and thus direct comparison would
not be appropriate.  Note that \cite{2020ApJ...898..105O} reported that
the \ion{Mg}{2} flux of ULAS J1342 was enhanced by $47^{+11}_{-11}$\%
when the \ion{Fe}{2} template of \cite{2001ApJS..134....1V} was used
instead of \cite{2006ApJ...650...57T}.  This is because the purely
empirical approach taken by \cite{2001ApJS..134....1V}, makes it
difficult to accurately estimate the contribution of \ion{Fe}{2} in the
wavelength range blended with \ion{Mg}{2}, and {\color{black} it}
significantly underestimates the \ion{Fe}{2} flux around \ion{Mg}{2}
compared to the semi-empirical \ion{Fe}{2} template of
\cite{2006ApJ...650...57T} obtained through photoionization simulation.

Previously, based on the accumulation of \ion{Mg}{2}/\ion{Fe}{2} flux
ratio measurements of quasars at various redshifts, many researchers
(e.g., \citealt{2011ApJ...739...56D};
\citealt{2017ApJ...849...91M}; \citealt{2019ApJ...874...22S};
\citealt{2020ApJ...898..105O}; \citealt{2020ApJ...905...51S}) claimed
that, despite a large scatter in the measurements, there was apparently
no redshift evolution in \ion{Mg}{2}/\ion{Fe}{2} over a wide range of
$z\sim 0\textendash7.5$.  However, when the \ion{Mg}{2}/\ion{Fe}{2} flux
ratios are converted to more physically meaningful [Mg/Fe] abundance
ratios, we emphasize that there indeed appears a large difference in
[Mg/Fe] between ULAS J1342 ($\mathrm{[Mg/Fe]} = -1.11 \pm 0.12$) and
quasars at $z<3$ (the approximate range of $\mathrm{[Mg/Fe]}$ extends
from $-0.5$ to $+0.2$), and this difference exceeds the uncertainties
associated with \ion{Mg}{2}/\ion{Fe}{2} measurements and their
conversion to [Mg/Fe].

\subsection{Standard Chemical Evolution for Host Elliptical Galaxies} \label{sec:std_model}

In this section, we examine whether the heavy elements
in the quasar BLR originate from chemical evolution in the host
elliptical galaxy.  For this purpose, it is not necessary to consider
the possibility of heavy-element formation in the BLR, but
rather decouple that region and consider chemical evolution only in the 
surrounding core region of the host elliptical galaxy.  Since elliptical
galaxies form early with a very large merging rate, star formation
in the core region begins in a burst.  However, the gravitational
potential is so large that gas does not flow out of the core and will be
used again in the next formation of stars therein.

In general, it has been widely accepted that the global
trend of chemical evolution over galactic scales is well described in
terms of spatially coarse averaging of heavy-element abundances as a
function of time (e.g., see the reviews by
\citealt{2012ceg..book.....M,2021A&ARv..29....5M}).  In the case of
elliptical galaxies, such a so-called one-zone model applies over the
core scale of a few kpc.  However, if we look into chemical phenomena
occurring on small scales without averaging, there is a large spatial
inhomogeneity in [Fe/H] of the local gas due to the asymmetric explosion
of SNe (e.g., \citealt{2019MNRAS.484.3307M}), associated later with
mechanical and radiative feedbacks (e.g., \citealt{2008ApJ...679..925W};
\citealt{2010AIPC.1294..128O}) as well as inhomogeneous metal mixing and
dilution (e.g., \citealt{2018MNRAS.475.4378C}),
etc.\footnote{The possibility of deriving information
about the global chemical evolution of galaxies from the correlation
analysis of the chemical abundance dispersion of stars and the
interstellar medium is described in \cite{2018MNRAS.475.2236K}.}

In describing the chemical evolution of galaxies, the
inhomogeneity effect is remarkable in low-density environments.  
However, in high-density environments considered in
this section, the inhomogeneity effect is much more suppressed, so that
coarse averaging of heavy-element abundances obtained over the core
scale would trace the global trend of chemical evolution in the core.
In the following, with the above limitations of spatial averaging in
mind, we calculate the standard model of chemical evolution for
elliptical galaxies hosting quasars.

The heavy elements observed in the host elliptical galaxy are thought to
be the result of chemical evolution driven by repetitive cycles of
active formation of stars and explosions of SN.  The first stars are the
Pop~III stars. They form from metal-free gas by hydrogen cooling. The
massive stars among them explode as SNe ejecting the first heavy
elements into the interstellar medium (ISM).  Next, the ultra metal-poor
stars of Pop~II form, initially by hydrogen cooling, from gas with
almost no heavy elements. The massive stars among them explode as SNe,
further increasing the amount of heavy elements in the ISM.  When the
heavy-element cooling supersedes the hydrogen cooling in the ISM, the
formation of metal-poor Pop~II stars is promoted.  This changeover from
the hydrogen cooling to heavy-element cooling occurs at $Z/Z_\sun \sim
10^{-4}$ (\citealt{1977ApJ...211..638S}; \citealt{1980PASJ...32..229Y};
\citealt{2001MNRAS.328..969B}).

The gas density at star-forming sites is $n_\mathrm{H} \sim
0.1$--$10^{3}~\mathrm{cm}^{-3}$ from the ambient medium to molecular
clouds in galaxies, and the velocity dispersion of the gas is typically
$v\sim 1$--$10~\mathrm{km~s^{-1}}$ (e.g.,
\citealt{2006ARA&A..44..367S}).  The SN, wherever it explodes, sweeps
out the surrounding gas to form a dense, expanding shell, and the
heavy-element yields of SN ejecta are mixed into the shell (e.g.,
\citealt{1988ApJ...334..252C}; \citealt{1998ApJ...507L.135S}).  The
expansion of the shell is halted by the turbulent pressure of the
surrounding gas. Eventually, the shell breaks due to the
Helmholtz-Kelvin instability on the shell surface or by encountering
neighboring shells. Then, the high-density gas in the shell falls back,
and the spherical volume surrounded by the shell is filled with diluted
gas. In this way, heavy elements diffuse throughout space, gradually
concentrating towards the galaxy center and being fed into the BLR gas
illuminated by the central engine.

According to nucleosynthesis calculations, the $\alpha$-element and iron
yields of the ejecta of a core-collapse supernova (CCSN) in the mass
range of $10\textendash 100~M_\sun$ are largely independent of the
original metallicity of its progenitor, even if it had zero metals
(\citealt{2002ApJ...565..385U}; \citealt{2007ApJ...660..516T};
\citealt{2010ApJ...724..341H}). Furthermore, assuming the Salpeter IMF ($\phi(m) \propto
m^{-1.35}$), the IMF-weighted theoretical yields reproduce the observed
value of $[\alpha/\mathrm{Fe}] \sim +0.4$ for metal-poor Pop~II stars in
the solar neighborhood and in local galaxies (e.g.,
\citealt{2009ARA&A..47..371T}). However, we note that all stars in
the 10--100~$M_\odot$ mass range that are CCSNe progenitors do not
necessarily explode and contribute to chemical enrichment
(\citealt{2018ApJ...857...46I}). Stars that do not explode but become
direct black holes are generally distributed discretely in the $>$15~$M_\odot$
mass range (e.g., \citealt{2011ApJ...730...70O};
\citealt{2015ApJ...801...90P}; \citealt{2021ApJ...909..169K}). In the
case of exploding stars, the yield ratio $y_i/y_j$ does not depend much
on the progenitor mass. Therefore, the IMF-averaged yield ratio $\langle
[y_i/y_j] \rangle$ remains almost unchanged even when non-exploding
stars are taken into account, and does not affect the chemical evolution
of galaxies.

Early in chemical evolution, a large amount of iron starts to be
supplied by SNe Ia at $t - t_f \approx
t_\mathrm{Ia,min}$ (\citealt{1995MNRAS.277..945T};
\citealt{1996ApJ...462..266Y}), where $t$ is the age of the universe,
$t_f$ is the age at which a galaxy was formed and
Pop~III stars were born, and $t_\mathrm{Ia}$ is the lifetime of SN Ia
which is much longer than CCSN.  Since the significant
iron enrichment is delayed by $t_\mathrm{Ia}$ compared to
$\alpha$-elements from CCSNe, a break occurs in
[$\alpha$/Fe] at $t-t_f \approx\ t_\mathrm{Ia,min}$, and
[$\alpha$/Fe] slowly decreases towards low redshift.  Such a decreasing
trend depends on the delay time distribution (DTD) for SN Ia. The DTD is
usually expressed as a power-law function $f_D(t_\mathrm{Ia})\propto
{t_\mathrm{Ia}}^{s}$ 
($t_\mathrm{Ia,min}\approx$ 0.1 Gyr), where the slope index 
is negative and has been reported as $s = -1$ (\citealt{2008PASJ...60.1327T};
\citealt{2014ARA&A..52..107M}) or $s = -1.4$
(\citealt{2020ApJ...890..140S}; \citealt{2021ApJ...922...15}).
The $t_\mathrm{Ia}$-model is constrained by SNe Ia 
which appeared in low-redshift galaxies below $z \sim 1.5$ and is extrapolated 
to higher redshift in this paper. 
This extrapolation is not intended to assume that SN Ia progenitors at 
higher redshift are the same as those at lower redshift, but to remain 
consistent with the existence of a generally accepted cosmic clock, 
based on the result that the $t_\mathrm{Ia}$-model reproduces the 
$[\alpha/\mathrm{Fe}]$ break observed in 
various stellar systems with their expected rates of star formation 
(the Galactic bulge, the solar  neighborhood, and local galaxies; e.g., 
\citealt{2012ceg..book.....M,2021A&ARv..29....5M}).

Later in chemical evolution, at $t - t_f > \Delta +
t_\mathrm{Ia,max}$, where $\Delta$ ($\approx$ 1 Gyr) is
the effective duration of early star formation, no SNe Ia occur anymore,
and the mass loss from the intermediate-mass Pop~II stars, which were
born in large numbers early times, supplies a large amount of material
to an ISM that has been already depleted by active star formation in the
early stages of chemical evolution.  Since this material has a subsolar
metallicity, $\mathrm{[Fe/H]} < 0$, and the Pop~II abundance ratio of
$[\alpha/\mathrm{Fe}] \sim +0.4$, [Fe/H] starts to decrease at $t - t_f
\approx \Delta + t_\mathrm{Ia,max}$ and [$\alpha$/Fe]
starts to increase.  Thus, the chemical evolution model is characterized
by three redshifts: $z_f$, $z(t_\mathrm{Ia,min})$, and
$z(t_\mathrm{Ia,max})$.

Figure \ref{fig:stdmodel} shows the evolution of [Mg/Fe] and [Fe/H] as a
function of the age of the universe predicted by the standard chemical evolution models 
for the core region of an elliptical galaxy with $z_f=10$ for two DTD cases of $s=-1$
($t_\mathrm{Ia}=0.1\textendash 3~$Gyrs), and $s=-1.4$
($t_\mathrm{Ia}=0.1\textendash 8~$Gyrs).  Other parameter values in
common are taken from \cite{1998ApJ...507L.113Y}, with some updates.
Although the two standard models reproduce the lower-redshift quasar
data reasonably well for both [Mg/Fe] and [Fe/H] at $z<3$, it is clear
that the standard models significantly overpredict [Mg/Fe] and
underpredict [Fe/H] compared to the respective values for ULAS J1342 at
$z=7.54$. Therefore, the standard models fail to reproduce the
higher-redshift quasar data at $z>7$.

Here, we call attention to the fact that, during the
CCSN-dominated early phase until the onset of SN Ia, the [$\alpha$/Fe]
of the model is always given by the IMF-averaged CCSN [$\alpha$/Fe] and
remains constant (CCSN plateau) at $[\alpha/\mathrm{Fe}] \sim +0.4$ for
the Salpeter IMF, regardless of which chemical evolution model is used
among different star formation rates with or without merging/outflow.
The only difference among such different models is the value of [Fe/H]
at which the [$\alpha$/Fe] break occurs, and is irrelevant to examining
whether $[\mathrm{Mg/Fe}] = -1$ for ULAS J1342 originates from the
chemical evolution of host galaxies.  Therefore, the failure of
reproducing the quasar data at $z > 7$ in this section based on elliptical
galaxies has a generality that holds for all other types of galaxies.

It is still worth considering a variant model in which the SN Ia ejecta
predominate in the early galaxy.  This is the case 
where a galactic wind is generated at $t-t_f \approx t_\mathrm{GW}$, and all the
residual gas is expelled outside of the galaxy and the formation of new
stars stops.  However, the intermediate-mass stars formed earlier
continue to release their iron-poor envelopes by mass loss, and SNe Ia
eject the iron-rich yield into the galaxy.  Accordingly, [Fe/H] rises
and [Mg/Fe] drops, immediately after $t_\mathrm{GW}$.  We run this model
with $t_\mathrm{GW}=0.2$~Gyr adjusted to more or less explain
$\mathrm{[Mg/Fe]}\sim -1$ for ULAS J1342 at $z=7.54$. 
We note that the galactic wind model described here 
is constructed only to show the maximum effect possible of 
SN Ia ejecta.
As shown
in Figure \ref{fig:stdmodel}, not only does this extreme 
model underpredict
[Fe/H] at $z=7.54$, it cannot also reproduce any of the lower-redshift
quasar data at all.\footnote{According to
\cite{2003ApJ...590L..83T}, there is a variation of $\pm25$\% in the
$M$(Ni) produced by SN~Ia. Even if $M$(Ni) increases by 25\% within this
range, [Fe/H] and [Mg/Fe] change by only $+0.1$ dex and $-0.1$ dex,
respectively. Therefore, the red curve in Figure \ref{fig:stdmodel} is
almost unaffected.}

In conclusion, as far as SNe Ia are considered as the main source that
supplies a large amount of iron into the ISM, it is evident that, in the
framework of chemical evolution of host elliptical galaxies, there
exists no solution that can explain the higher-redshift quasar data at
$z>7$ and the lower-redshift quasar data at $z<3$ simultaneously for
both [Mg/Fe] and [Fe/H].


\section{Alternative Scenario of Pop~III Initiated Chemical Evolution}

The currently successful paradigm of galactic chemical evolution is that
the diversity of observed patterns of heavy-element abundances at
various redshifts for a certain type of galaxy would always be
reproduced by a properly selected redshift-dependent ratio of the
numbers of SN Ia to CCSN that reflects the star
formation history considered (\citealt{1995MNRAS.277..945T}).  In fact,
the lower-redshift data of [Mg/Fe] and [Fe/H] for quasars at $z<3$ are
well reproduced by the standard chemical evolution models of host
elliptical galaxies. However, the value of [Mg/Fe]$\sim -1$ for ULAS
J1342 at $z=7.54$ is far too low to be consistent with these models
under this paradigm.

\begin{figure}[t]
 \epsscale{1.1}
 \plotone{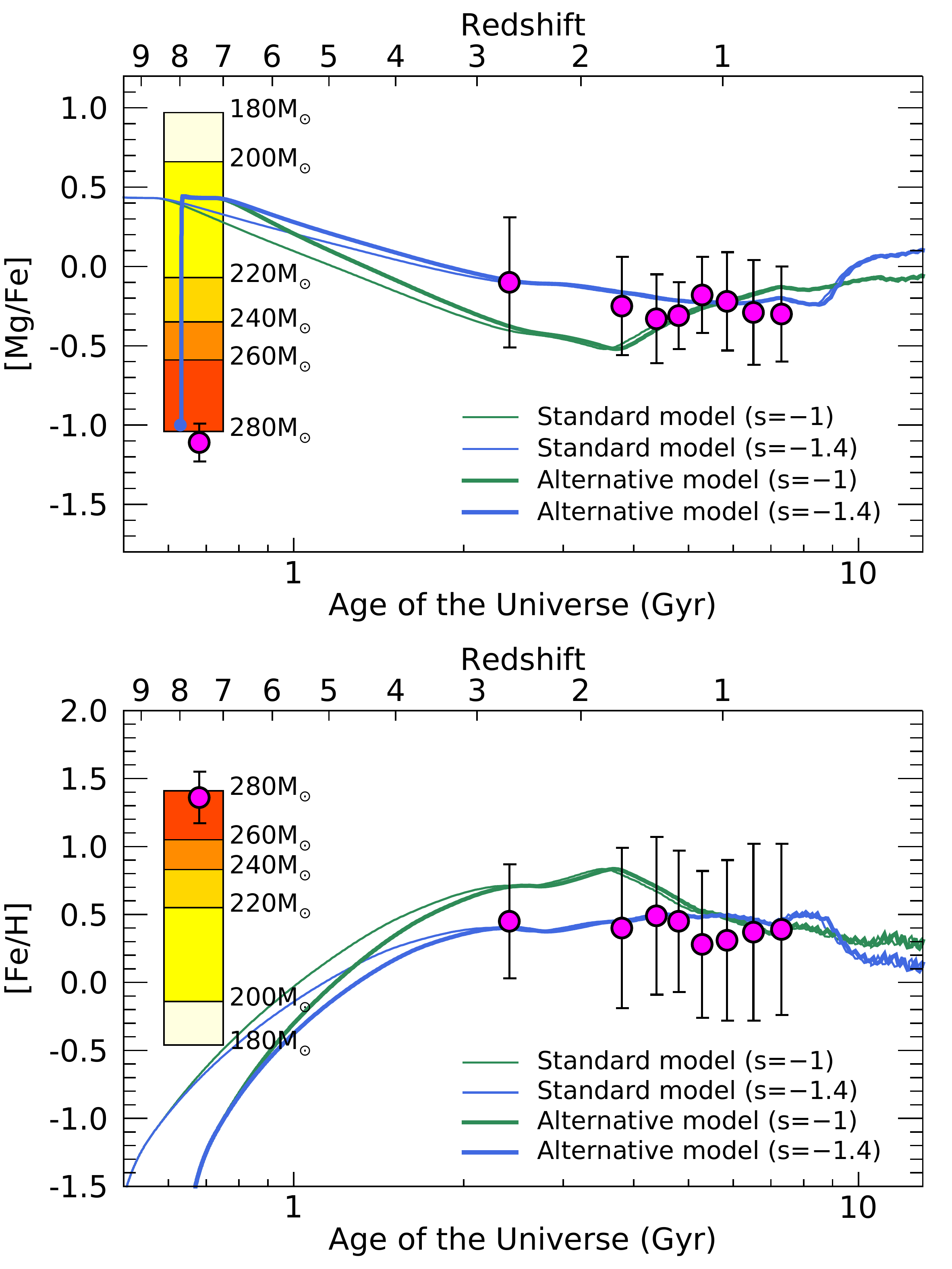}
 \caption{[Mg/Fe] (upper panel) and [Fe/H] (lower panel) as a function
 of the age of the universe.  Same as Figure~\ref{fig:stdmodel}, 
 but the alternative chemical evolution models for the core region of 
 an elliptical galaxy, which is thought to host ULAS J1342, are shown 
 by thick curves, in addition to the standard models. 
 The alternative models are calculated by assuming 
 that a fraction of the BLR gas with [Fe/H]$= +1.3$ and [Mg/Fe]$=-1$ 
 is initially mixed with the metal-free gas in the core region of the 
 elliptical galaxy, and that the chemical evolution in the core region 
 starts with [Fe/H]$= -4$ and [Mg/Fe]$= -1$ at $z=8$ 
 (see the text in \S{3}).
 The color-coded rectangular regions
 indicate the values of [Mg/Fe] and [Fe/H] as a function of progenitor
 mass of Pop~III PISN.  The width of these regions corresponds to the
 transition period of cosmological reionization.
 }
 \label{fig:pop3model}
\end{figure}

Such an unusually low value of [Mg/Fe] can only be explained by an SN
that ejects a much larger iron yield than CCSN.  Given
that SN Ia was shown to be an unlikely candidate (see
\S\ref{sec:std_model}), the most promising candidate is the
pair-instability supernova (PISN) caused by the explosion of massive
Pop~III stars in the mass range of $150\textendash 300~M_\sun$
(\citealt{2002ApJ...565..385U}; \citealt{2002ApJ...567..532H}).  In
particular, based on recent nucleosynthesis calculations
(\citealt{2018ApJ...857..111T}), the abundance ratio of [Mg/Fe], as a
function of the progenitor mass, is indicated by the color-coded
rectangular region shown on the upper panel of Figure
\ref{fig:pop3model}.  The horizontal width of this region corresponds to
the transition period of cosmic reionization reported to have
effectively occurred at $z \sim 7.7\pm 0.7$ (e.g.,
\citealt{2020A&A...641A...6P}).  It is evident from this figure that
[Mg/Fe]$\sim -1$ is the minimum value for the Pop~III PISN ejecta, and
is not achievable by any other PISNe except the one with
$280~M_\sun$\footnote{The Pop~III PISN ejecta of rotating progenitors
are also calculated by \cite{2018ApJ...857..111T}; given the mass of
progenitor, the value of [Mg/Fe] for the case of rotation is smaller by
about $-0.2$ dex than that without rotation.}.  Accordingly, we deduce
the PISN near the high-mass end as the iron source to the BLR gas of
ULAS J1342 at $z=7.54$.

The iron abundance of ULAS J1342 has been estimated to be
$[\mathrm{Fe/H}]=+1.36$.  On the other hand, the iron yield ejected by a
Pop~III PISN with $280~M_\sun$ is about $40~M_\sun$, and the [Fe/H] of
the SN ejecta is several hundred times greater than the solar value.
When this super metal-rich ejecta is mixed into the metal-free BLR gas,
the BLR mass required for ULAS J1342 to give $[\mathrm{Fe/H}] \sim +1$
is about $10^3~M_\sun$.  Then, assuming a BLR mass of $10^3~M_\sun$,
[Fe/H] in the BLR monotonically decreases as the PISN progenitor mass
decreases below $280~M_\sun$.  This trend is indicated by the
color-coded rectangular region shown on the lower panel of
Figure~\ref{fig:pop3model}.

The BLR mass has been estimated so far by several other methods (e.g.,
\citealt{1989agna.book.....O,1997iagn.book.....P,1999agnc.book.....K}).
Their results are very different from each other, and the BLR mass is
still a poorly understood quantity.  In \cite{2003ApJ...582..590B}, the
BLR is considered to have multiple components, and its mass depends on
the monochromatic luminosity at 1450~\AA\ ($\lambda L_{1450}$).
Inserting $\lambda L_{1450}$ measured from the spectrum of ULAS
J1342\footnote{Since \cite{2020ApJ...898..105O} did not provide a value
for $\lambda L_{1450}$, we roughly estimated $\lambda L_{1450} \sim 2
\times 10^{44}~\mathrm{erg~s^{-1}}$ from their spectrum in Figure 1 by
eye.}  into their formula gives a BLR mass in the revised range of
$10^3\textendash 10^4~M_\sun$, consistent with our estimate of
$10^3~M_\sun$.

We note that the gas density and velocity dispersion of the BLR are
$n_\mathrm{H} \sim 10^{10-12}~\mathrm{cm}^{-3}$ (e.g.,
\citealt{1992ApJ...387...95F}; \citealt{2017ApJ...834..203S}) and $v\sim
10^{3-4}~\mathrm{km~s^{-1}}$ (e.g., \citealt{2006LNP...693...77P};
\citealt{2011ApJS..194...45S}), respectively.  These values are many
orders of magnitude greater than those at the star-forming sites in
galaxies, making it impossible for a new star to be born in the
BLR. Instead, it is conceivable that a massive Pop~III star, which
eventually explodes as Pop~III PISN, would be formed synchronously with
the processes by which the broadly distributed metal-free gas
accumulates to create a massive black hole (BH) at the very
center of a protogalactic clump.

In this situation, the explosion of Pop~III PISN in the
extremely dense gas, which is later illuminated by the
central engine to become a BLR, is the first and last Pop~III event 
of the quasar to which this BLR belongs. Once it explodes, the 
heavy-element yields from this SN accumulate in the shocked gas  
around the SN (e.g., \citealt{1992MNRAS...255...713T}; 
\citealt{2019MNRAS.488.978J}), and are immediately mixed 
with the entire metal-free BLR gas, through the destruction of 
shock front due to hydrodynamical instabilities in the late 
phase of its expansion. 
Subsequently, it is expected that large-scale mixing of the 
metal-enriched BLR gas would collectively occur 
over a scale of protogalactic clumps that grow by 
gravitational aggregation. When the average value of the 
amount of heavy elements in such clumps is raised 
to $Z/Z_\sun \sim 10^{-4}$, the heavy-element cooling  supersedes the 
hydrogen cooling and starts to effectively form extremely metal-poor 
Pop~II stars in sufficiently cool and quiescent regions. On the other 
hand, the abundance ratio of [Mg/Fe] in the BLR is 
preserved during the process of large-scale mixing, 
because  this mixing would not be considered as proceeding differently 
among different heavy elements. 

As a rough sketch of our alternative scenario leading 
to the chemical evolution in the inner core region of an elliptical galaxy, 
which is thought to host ULAS J1342, we assume that a massive Pop III 
star with 280Ms forms first 
in the central region of a protogalactic clump of dense gas, and explodes at 
the end of lifetime of 2~Myr as PISN (\citealt{2002A&Ap...382...28}). 
Its heavy-element ejecta with $[\mathrm{Mg/Fe}] \sim -1$ 
(\citealt{2018ApJ...857..111T}) is mixed throughout 
the surrounding dense gas. Strong UV radiation from the central engine 
is then generated by the gas infall which grows the central BH through 
the accretion disk, and illuminates the surrounding dense gas which 
thereafter becomes a BLR. 
The mass decrease in the BLR by the infall is 
replenished by mass supply from the inner core that 
surrounds the BLR from outside. The cycle of mass decrease 
and supply keeps the central engine running as long as the infall continues.

At the same time, a part of the gas infall 
is reversed to drive the outflow from active 
galactic nucleus (AGN) (e.g., \citealt{2021NatRep...5...13}).
This outflow as well as others by large-scale mixing is contaminated in the 
reservoir of a host galaxy, and some fraction of the gas therein is expected 
to move inwards by angular momentum transport to the scale of inner core 
of a host galaxy (e.g., \citealt{2020ARA&A...58...27}). 
Such a gas circulation increases the iron abundance in the core gas above 
$[\mathrm{Fe/H}] = -4$ to promote the efficient formation 
of Pop~II stars.

After such a transition to Pop~II star 
formation above $[\mathrm{Fe/H}] = -4$, 
PISN events only rarely occur, if any at all, and numerous 
Pop II CCSNe dominate. Therefore, the core gas 
is enriched by CCSNe with $[\mathrm{Mg/Fe}] \sim +0.4$,  
and the inflow of this gas into the BLR makes the 
nucleosynthetic signature of the Pop~III PISN dissapear. 
In other words, the visibility time, such that the 
Pop~III PISN $280~M_\sun$ signature with $[\mathrm{Mg/Fe}] \sim -1$ 
remains dominant since its explosion in the BLR of ULAS J1342, corresponds to 
at least the lifetime of the stars in the higher-mass portion of 
the IMF among all CCSN progenitors with $[\mathrm{Fe/H}] = -4$, 
which is about 3 Myr for $50\textendash 100~M_\sun$ 
(\citealt{2002A&Ap...382...28}). We will discuss the consequence 
of this visibility time in \S4. 

Based on the above consideration, Figure \ref{fig:pop3model} shows the
evolution of [Mg/Fe] and [Fe/H] as a function of the age of the
universe, predicted by the alternative chemical evolution models 
for the core region of an elliptical galaxy for the
two DTD cases of $s=-1$ and $-1.4$, but for the initial conditions at
$z=8$ set to $\mathrm{[Fe/H]} = -4$ as given and $\mathrm{[Mg/Fe]} =
-1$ as measured for the BLR of ULAS J1342.  For the purpose of
comparison, the standard models in Figure \ref{fig:stdmodel} are also
shown in this figure.  Apparently, the alternative models begin
with a sudden rise in [Mg/Fe], immediately reaching the Pop~II
CCSN plateau level of $[\mathrm{Mg/Fe}] \sim +0.4$, and
then slowly decline towards a low redshift.  Moreover, when compared
with the standard models with $z_f=10$, the alternative models begin 
with a steeper rise in [Fe/H] starting from [Fe/H]=-4 at $z=8$.

Alternative models beginning with different combinations of [Mg/Fe] and
$z$ are also possible, following the Pop~III PISN explosion of massive
stars that form below $280M_\sun$ at different redshifts in the
transition period of cosmic reionization. Therefore, future observations
of quasars at higher redshifts exceeding $z\sim 7$ are expected to show
that such quasars fill in the rectangular regions, as illustrated in the
upper and lower panels of Figure~\ref{fig:pop3model}.


\section{Discussion} \label{sec:discussion}

In this paper, we deduced the origin of $[\mathrm{Mg/Fe}] \sim -1$ in
ULAS J1342 as a Pop~III PISN with $280~M_\sun$.  According to recent
nucleosynthesis calculations (\citealt{2018ApJ...857..111T}), the ejecta
of this PISN has the lowest [Mg/Fe] value in the mass range of PISN
progenitor stars.  On the less massive side below $280~M_\sun$, [Mg/Fe]
increases monotonically as the mass of PISN progenitor stars decreases.
On the more massive side above $280~M_\sun$, [Mg/Fe] becomes much
larger, beyond that of a PISN explosion due to the significantly smaller
ejecta of iron from CCSN progenitor stars in this mass range.
Therefore, in a sample of $z > 7$ quasars around $[\mathrm{Mg/Fe}] =
-1$, the mass distribution of Pop~III stars can be derived from the
observed [Mg/Fe] distribution of quasars, using the relationship between
[Mg/Fe] of the ejecta and progenitor mass of Pop~III PISNe.

According to 3D cosmological simulations over $3~\mathrm{Mpc}^3$ by
\cite{2015MNRAS.448..568H}, the predicted mass distribution of Pop~III
stars formed predominately by $\mathrm{H}_2$ cooling below $z \sim 30$
has a peak around
$200\textendash300~M_\sun$.\footnote{The mass
distribution of Pop~III stars born from the primordial gas also has a
peak at mass $15\textendash 40~M_\odot$, which is produced by HD cooling
(\citealt{2015MNRAS.448..568H}; see also \citealt{2018MNRAS.475.4378C}).
Pop~III stars belonging to this peak are CCSN progenitors, and the
ejecta is $[\mathrm{Mg/Fe}] > 0$, thus $[\mathrm{Mg/Fe}] = -1$ for ULAS
J1342 cannot be explained.}  The location of this peak is 
not very
dependent on the range of redshift in which Pop~III stars were formed,
nor on whether the effect of external UV radiation during the formation
of Pop~III stars is taken into account in the simulations.  Note that
the mass distribution has a shape with an asymmetrical decline away from
the peak; on the massive side, it declines more rapidly than on the less
massive side. To compare future observations with cosmological
simulations, we need to carefully examine various biases in assessing
the rate of detecting high-redshift quasars and the rate of hosting
Pop~III stars in protogalactic clumps.  However, if future quasar
observations reveal a sign of the peaked mass distribution of Pop~III
stars, it will certainly impose constraints on the modeling of structure
formation in the universe.

If the heavy elements in quasars at $z > 7$ originated from Pop~III
PISNe, the quasars in the $\mathrm{[Mg/Fe]}\textendash z$ diagram are
expected to be distributed in the vertical direction of the rectangular
region from $[\mathrm{Mg/Fe}] \sim -1$ to $+1$, reflecting the mass
range of the PISN progenitor stars.  On the other hand, cosmological
simulations show that not all primordial halos of $M_\mathrm{halo} <
10^8~M_\sun$ accompany the formation of massive Pop~III stars of
$10^{2\textendash3}~M_\sun$ (e.g., \citealt{2015MNRAS.448..568H},
\citealt{2016ApJ...823..140X}).  Those halos without Pop~III stars
assemble into protogalactic clumps that undergo the chemical evolution
driven by the cycles of formation of stars and explosion of CCSN and SN Ia.  
Then, the quasars, triggered to turn
on in the central region of host elliptical galaxies, are distributed
like a ridge along the standard chemical evolution models, from which
the [Mg/Fe] break can be constrained
(\citealt{2017ApJ...834..203S,2020ApJ...904..162S}).

It is interesting to note that, for a fairly large sample of quasars at
$z = 6\textendash 8$, we expect to see a distinctive cross-shaped
feature in the $\mathrm{[Mg/Fe]}\textendash z$ diagram where two
different quasar distributions intersect with each other at right
angles.  The vertical and horizontal distributions correspond to quasars
for which heavy elements originate from Pop~III PISNe and host
elliptical galaxies, respectively.  If this is confirmed, the [Mg/Fe]
break will be buried at the intersection of the two distributions,
making it difficult to accurately determine the redshift of the [Mg/Fe]
break without separating the quasars that belong to different
distributions.  This separation is a future issue, but it seems possible
by measuring the abundance ratio of heavy elements other than [Mg/Fe].

The reliability of the idea that heavy elements in quasars originate
from two sources of Pop~III PISNe and host elliptical galaxies depends
on whether a large-scale mixing of Pop~III ejecta can increase the
average heavy-element abundance of primordial gas to $Z/Z_\sun =
10^{-4}$, which activates star formation and promotes chemical evolution
in protogalactic clumps distributed over a scale of Mpc in the universe.

The comoving baryon density of the Big Bang cosmology can 
be expressed
as $\rho_b = \Omega_b h^2 \times 2.5 \times
10^{11}~M_\sun/\mathrm{Mpc}^3$, where $\Omega_b$ is constrained from the
Big Bang nucleosynthesis and given by $\Omega_b h^2 = 0.022$ from
observations of light elements (\citealt{2000IAUS..198..125T}) and the
CMB (\citealt{2020A&A...641A...6P}).  On the other hand, if heavy
elements in ULAS J1342 during cosmic reionization originate from a
Pop~III PISN with $280~M_\sun$, the iron yield of its ejecta is about
$40~M_\sun$, which can raise the metallicity to $Z/Z_\sun = 10^{-4}$
when mixed with a primordial gas of $4 \times 10^8 M_\sun$.  In other
words, the number of such SNe required to make $Z/Z_\sun =10^{-4}$ for a
primordial gas of $1~\mathrm{Mpc}^3$ cube is only about 20, and the
average separation between SNe is about 400~kpc.  Note that the required
number of PISNe is much larger if a Salpeter-like mass function is
assumed and more weight is given on the smaller iron yield from PISNe
below $280~M_\sun$ (e.g., \citealt{2005MNRAS.360..447M}).

The above argument assumes that a large-scale mixing of heavy elements
over a scale of Mpc would operate to achieve a homogeneous metallicity
distribution in the universe.  However, this scale is not reached by
conceivable processes such as a hydrodynamical mixing associated with
gravitational aggregation into protogalactic clumps.  Therefore, in
reality, the metal enrichment beyond $Z/Z_\sun=10^{-4}$ is localized in
space, and the metallicity distribution in the universe becomes
necessarily inhomogeneous.  According to cosmological simulations, the
degree of inhomogeneity is measured by the fraction of the volume in the
universe occupied by gas of pristine composition.  This fraction is
estimated to exceed 80\% at $z=7.6$ (\citealt{2016ApJ...823..140X}; see
also the review by \citealt{2018FrASS...5...34N}), providing the
possibility for the formation of Pop~III stars in a significant range of
redshifts around $z = 7.6$.

According to explosive nucleosynthesis calculations, the [Mg/Fe] 
ratio used to deduce a Pop III PISN with $280~M_\odot$ in this paper does 
depend on the progenitor mass. However, various other abundance ratios have 
been proposed that do not depend much on the progenitor mass. 
When measured by the flux-to-abundance conversion described in 
this paper, such other ratios in the BLR are expected to show a 
Pop III PISN-characteristic, such as a high ratio of [Si/Mg]$\sim +1$ for the 
large Si excess and a low ratio of [Al/Mg]$\sim -1$ for the odd-even effect 
(\citealt{2018ApJ...857..111T}).  Furthermore, PISN models with 
the initial metallicity of $Z=0.001$ show an $\alpha$-element abundance 
pattern similar to Pop III PISN, but with the significantly weaker odd-even effect 
(\citealt{2014A&A...566...A146K}). The emission lines of \ion{Si}{4} $\lambda1397$ 
and \ion{Al}{3} $\lambda1857$ are not far from the \ion{Mg}{2} $\lambda2798$ and the 
UV \ion{Fe}{2} bump in $2200\textendash3090\AA$. Therefore, all these lines are 
simultaneously observed in the near-infrared spectrum of quasars at $z > 7$, 
avoiding systematic errors due to the use of different instruments and telescopes, 
so that the [Si/Mg] and [Al/Mg] ratios for the BLR can be measured with high reliability 
and will hopefully agree with those theoretically predicted for Pop III PISNe. 
The above follow-up studies should strongly confirm the 
existence of a Pop III PISN not only with $280M_\odot$ in particular, but also over 
the range of $150-300~M_\odot$ in general.

The PISN event rate (ER) is proportional to a product of visibility 
time and PISN occurrence rate density (e.g., \citealt{2005ApJ...633...1031}). 
In \S{3}, we estimate that the visibility time for the nucleosynthetic PISN 
signature in the BLR of ULAS J1342 is at least 3 Myr, which is about six orders of 
magnitude longer than the visibility time for the photometric PISN signature 
(about 1yr in the source frame), based on the theoretical light curve immediately 
after the explosion of PISN (\citealt{2011ApJ...734...102K};  
\citealt{2014ApJ...797..9W}; \citealt{2018MNRAS.479.2202H}). 
On the other hand, the PISN occurrence rate density for a population of quasars 
or galaxies that do not harbor a quasar at the galaxy center, is assumed to be 
proportional to the number density of each population if the survey area is common. 
Using the normalisation factor $\Phi^*$ of the luminosity function (LF), 
we tentatively obtain $\Phi^*$(quasar)/$\Phi^*$(galaxy)
$\sim 10^{-4}$ at $z = 7$  
(quasar LF: \citealt{2018ApJ...869...150}; 
galaxy LF: \citealt{2015ApJ...803...34}), 
though this value is possibly underestimated, because the observation of quasar LF 
at $z=7$ is still ongoing. Thus, the PISN ER, for which the candidate sites of 
PISNe have already been targeted to quasars, is at least a few hundred times 
larger than that for the blind deep survey of galaxies. As a result, our finding 
in this paper of a Pop~III PISN signature in a high-redshift quasar of ULAS J1342 
is not unlikely.

The existence of a massive Pop~III star at high
redshift, if confirmed, may impact ongoing searches for signatures of
massive Pop~III stars imprinted on the heavy-element abundance patterns
of second-generation stars in the Galaxy.  Current search strategies
encounter difficulties in how to identify true second-generation stars
from a sample of very metal-poor (VMP; $[\mathrm{Fe/H}] < -2]$) or EMP
($[\mathrm{Fe/H}] < -3]$) stars. The correlation between the metallicity
of a star and the age of the Galaxy at its birth is not considered to
hold as the metallicity decreases below [Fe/H] $\sim -2.5$
(\citealt{1998ApJ...507L.135S}).  Since selecting EMP stars does not
necessarily mean dating back in time to the second-generation stars, the
nucleosynthetic features of massive Pop~III stars may not be constrained
from the heavy-element abundance patterns of the such stars.  However,
if true second-generation stars can indeed be selected, the type of
Pop~III SNe responsible for their observed abundance patterns can be
specified.  Then, the IMF of the Pop~III progenitors can be derived from
the variety of these patterns, which depend on the progenitor mass of
Pop~III SNe.

If the result of Pop~III PISN in this paper is applicable to the Milky
Way, the [Mg/Fe] abundance ratios of second-generation stars range over
$-1 \lesssim $ [Mg/Fe] $ \lesssim +1$, depending on the progenitor mass
of PISNe in the range of $150\textendash300~M_\sun$ 
(\citealt{2018ApJ...857..111T}).  On the other hand, their [Fe/H] values
would be distributed in some range, depending on how much the super
metal-rich iron yield of PISN ejecta was diluted by being mixed with gas
of pristine composition before a favorable environment for formation of
second-generation stars was achieved.  Their range of [Fe/H] can then be
estimated from the VMP stars that have much smaller values of [Mg/Fe]
than the observed average of $+0.3 \lesssim $ [Mg/Fe] $\lesssim +0.5$
for the VMP stars.  In this way, a sample of candidate second-generation
stars may be obtained by selecting the stars in the estimated range of
[Fe/H], and further separating them into the PISN or CCSN origin
according to their theoretical heavy-element abundance patterns.  

In a previous result along this line, \cite{2014Sci...345..912A} reported
that the VMP star SDSS J0018-0939 has a very low [C/Fe], as well as
significantly low [Mg/Fe] and [Co/Fe] ratios, and is best fit to the
theoretical PISN abundance pattern. At present,
this is the only one reported case of a star in the Galaxy showing 
signs of PISN, but this may be because there is no positive motivation to 
consider a star with an observationally low [Mg/Fe] as 
a second-generation star. 
New-generation photometric surveys,
such as J-PLUS (\citealt{2019A&A...622A.176C}) and S-PLUS
(\citealt{2019MNRAS.489..241M}; \citealt{2021ApJ...912..147W}), which
employ multiple narrow-band filters (including filters centered on C and
Mg features) in combination with wider-band filters, are currently
underway, and have the potential to efficiently identify stars with 
very low [Mg/Fe] at all metallicities.  Eventual derivation of the
Pop~III IMF from the [Mg/Fe]-selected PISN sample of VMP/EMP stars
discovered in these surveys will certainly be a future challenge for
Galactic astronomy.


\section{Caveats and Alternative Interpretations}

\subsection{Pop~III CCSN with $1000~M_\sun$}

In order to better constrain the origin of $[\mathrm{Mg/Fe}] \sim -1$ in ULAS
J1342, it is worth considering other sources  
to account for significant iron
ejection from Pop~III SNe.  The upper limit of the mass range for first
stars is uncertain, but the possibility of $500\textendash 1000~M_\sun$
has been pointed out by a number of authors. Explosions of the very massive stars in this
range require a collapsar-powered engine.  Assuming
the jet energy injection from the accretion disk, and setting the jet
parameters to cause the star to explode successfully,
\cite{2006ApJ...645.1352O} performed explosive nucleosynthesis
calculations of Pop~III CCSNe with 500~$M_\odot$ and 1,000~$M_\odot$.
While the nucleosynthesis of such a collapsar model is highly uncertain,
it is worth mentioning that their 1,000~$M_\odot$ model (A-1), among
others, interestingly gives $[\mathrm{Mg/Fe}]=-1.17$, which explains the
$[\mathrm{Mg/Fe}]=-1.11 \pm 0.12$ of ULAS J1342.  However, the BLR mass
that can explain $[\mathrm{Fe/H}]\sim +1$ is about $300~M_\sun$, which is
even smaller than the progenitor mass, and far below the acceptable range
by \cite{2003ApJ...582..590B}, and therefore unlikely.  In addition, the
$3~\mathrm{Mpc}^3$ cosmological simulations by
\cite{2015MNRAS.448..568H} identified the formation of 1540 primordial
gas clouds, of which only two Pop~III stars larger than $1000~M_\sun$
were formed and associated with such clouds.  It is unlikely that this
extremely rare case occurred in ULAS J1342.

\begin{deluxetable}{lcc}[t]
\tablecaption{Nucleosynthesis of Pop~III SNe \label{tab:pop3_yield}}
\tablehead{
 \colhead{Abundance ratio} & \colhead{PISN ($280~M_\sun$)} &
 \colhead{CCSN ($1000~M_\sun$)}
}
\startdata
$[\mathrm{Mg/Fe}]$     & $-1.06$ & $-1.19$ \\
$[\mathrm{Si/Fe}]$     & $-0.08$ & $-1.18$ \\
$[\mathrm{O/Fe}]$      & $-0.84$ & $-1.30$ \\
$[\mathrm{(Si+O)/Fe}]$ & $-0.73$ & $-1.29$ \\
\enddata
\tablerefs{\cite{2018ApJ...857..111T} for PISN;
 \cite{2006ApJ...645.1352O} for CCSN.}
\end{deluxetable}

A more direct distinction between Pop~III PISN with $280~M_\sun$ and Pop
III CCSN with $1000~M_\sun$ (A-1) can be made by measuring [Si/Fe] using
Si emission lines.  In practice, since \ion{Si}{4} $\lambda1397$, the
strongest Si emission line in the UV wavelength of a quasar spectrum, is
blended with [\ion{O}{4}] $\lambda1402$, only an upper limit of
\ion{Si}{4} $\lambda1397$ flux can be obtained.  Therefore, it is more
realistic to measure the blended line flux of \ion{Si}{4}+[\ion{O}{4}].
As shown in Table~\ref{tab:pop3_yield}, even in this case, the
difference in [(Si+O)/Fe] between Pop~III PISNe with $280~M_\sun$ and
Pop~III CCSNe with $1000~M_\sun$ based on nucleosynthesis calculations
is large enough to be distinguished by observation.
EW(\ion{Si}{4}+[\ion{O}{4}]) can be measured from the NIR spectrum of
ULAS J1342, and [(Si+O)/Fe] can be derived by constructing the
\ion{Si}{4}+[\ion{O}{4}] flux to the abundance conversion grid using our
method (\S\ref{sec:measure}).  It is important to emphasize that our
result that the heavy elements in ULAS J1342 are of Pop~III SN origin
remains unchanged irrespective of either candidate.

\subsection{A SNe Ia Cluster Model}

In \S2.3, we excluded SNe Ia, a recognized major source of 
iron, from the candidates explaining the 
observation of such an unusually low ratio of 
$[\mathrm{Mg/Fe}]\sim -1$ in the BLR of ULAS J1342 
at $z=7.54$. This exclusion was made because the iron 
in the BLR was assumed to originate from 
the early chemical evolution of an elliptical galaxy 
hosting ULAS J1342, and even the maximum possible iron 
contribution from CCSN + SN Ia was found unable to 
reproduce the observation of $[\mathrm{Mg/Fe}]$ 
(see Fig.3). However, if the iron in the BLR did not 
originate from the host galaxy, but solely from SNe Ia 
in the BLR, it is still worth considering a SN Ia with 
theoretical ejecta of $[\mathrm{Mg/Fe}]\sim -1.5$  
(\citealt{1997NucPhysA...621...467};
\citealt{1999ApJS...125...439}). 

In fact, it has long been pointed out that a white dwarf 
(WD) of near-Chandrasekhar mass is caused to explode as a
SN Ia by the tidal disruption when passing close to a BH 
(\citealt{1989A&A.209.103}; \citealt{2004ApJ.610.368}). 
Should this phenomenon actually occur near the massive 
BH at the center of a high-redshift galaxy, the BLR might 
be a unique site exhibiting the nucleosynthetic features 
solely from SNe Ia. On the other hand, 
the iron yield from one SN Ia is $\sim 0.5~M_\sun$, 
which is only about one hundredth of the iron yield from  
one PISN with $280~M_\sun$. Moreover, the BLR mass 
required by one SN Ia to explain the observation of 
$[\mathrm{Fe/H}]\sim +1$ by dilution is also only about 
one-hundredth of the BLR mass estimated in this paper (\S3).
Consequently, a cluster consisting of 100 SNe Ia in the 
BLR, though highly speculative, would lead to more or less 
the same result as one PISN with $280~M_\sun$.

The problem of distinguishing such a SNe Ia cluster from 
a PISN can be solved by comparing their heavy-element 
abundance ratios (other than [Mg/Fe]), notably 
$[\mathrm{Al/Mg}]$, which exhibits an odd-even effect.  
Nucleosynthesis calculations give 
$[\mathrm{Al/Mg}]\sim -1$ for Pop~III PISN with almost 
no dependence on progenitor mass
(\citealt{2018ApJ...857..111T}), while 
$[\mathrm{Al/Mg}]\sim -0.2$ for SNe Ia, with almost no 
dependence on the WD model (\citealt{1999ApJS...125...439}).
It is therefore expected that this significant difference 
in $[\mathrm{Al/Mg}]$ could provide a clear distinction 
between SNe Ia and a PISN. In the near-infrared spectrum of 
ULAS J1342 at $z=7.54$, the emission lines of \ion{Al}{3} 
$\lambda1857$ and \ion{Mg}{2} $\lambda2798$ are both 
visible, and their EWs can be measured. However, an idea 
of SNe Ia cluster remains speculative, as 
$[\mathrm{Al/Mg}]$ depends on it being reliably measured by converting the 
heavy-element EWs to their abundances by our method 
(\S\ref{sec:measure}).

\subsection{A Single Stellar Population (SSP) Model}

The formation and early evolution of galaxies have usually 
been discussed from various aspects, based on the so-called 
single stellar population (SSP) model (e.g., 
\citealt{2002A&Ap...382...28}; 
\citealt{2018MNRAS.479.2202H}). 
In this model, Pop III stars are assumed to form 
instantly on short timescales according to a log flat or 
Salpeter-like IMF, and Pop III stars end their life 
as SNe sequentially from massive to less-massive 
stars. When the IMF is extended to the mass range of PISN 
progenitor stars ($150\textendash 300~M_\sun$), they all 
end their life at once because their lifetime is 
estimated as 2 Myr, independent of their progenitor mass
(\citealt{2002A&Ap...382...28}). In this case, 
different amounts of heavy-element yields from PISNe with 
different progenitor masses are all mixed together in the 
gas, so that the heavy-element abundance pattern 
characteristic of PISNe at the high-mass end cannot dominate, 
contrary to our interpretation of the observation of 
$[\mathrm{Mg/Fe}]\sim -1$ based on a Pop III PISN with 
$280~M_\sun$. 

Currently, there has been no confirmation that the SSP 
model is applicable to the BLR gas, nor, in the first 
place, that stars can actually form from the BLR gas having 
extreme density ($10^{10-12}~\mathrm{cm}^{-3}$) and velocity 
dispersion ($10^{3-4}~\mathrm{km~s^{-1}}$), which are 
completely different from normal star-forming sites 
in galaxies. On the other hand, it has been reported from 
cosmological simulations that the rapid gas infall towards 
the center of a primordial star-forming cloud and the 
accretion onto the central stellar core form one massive 
Pop III star per each star-forming cloud 
(\citealt{2015MNRAS.448..568H}).  According to this study, 
the mass distribution for an ensemble of massive Pop III 
stars in the universe has a peak around the high-mass end 
of Pop III PISN, consistent with our interpretation of 
$[\mathrm{Mg/Fe}]\sim -1$. 
If massive Pop III stars with $\gtrsim 100~M_\sun$ form only via such specific 
processes, detection strategies so far for PISNe may 
necessarily be affected. Research along this line is required 
in the future.


\section{Summary and Conclusion}

Using the NIR spectrum of the quasar ULAS J1342 at $z=7.54$, we have
estimated the chemical abundance at BLR to be $[\mathrm{Mg/Fe}] = -1.11
\pm 0.12$ and $[\mathrm{Fe/H}] = 1.36 \pm 0.19$.  This paper is
summarized as follows:

\begin{enumerate}
 \item Our abundance diagnostic method requires correction for the
       luminosity dependence of the EW, but there is still uncertainty
       as to how the fiducial luminosity should be set at $z > 7$.  In
       this study, we extrapolated the results of
       \cite{2020MNRAS.495.3252S}, which give characteristic
       luminosities up to $z=7$, and applied them to ULAS J1342.
 \item The estimated value of $[\mathrm{Mg/Fe}] \sim -1$ for ULAS J1342
       was found to be significantly different from $z < 3$, which
       exceeds the uncertainties associated with \ion{Mg}{2}/\ion{Fe}{2}
       measurements and their conversion to [Mg/Fe].  ULAS J1342 seems
       to be richer in iron than quasars at $z < 3$.
 \item The standard model in the core region of an
       elliptical galaxy hosting a quasar, which takes into account SN
       Ia and CCSN, cannot explain [Mg/Fe] and [Fe/H]
       of ULAS J1342.  This conclusion based on
       elliptical galaxies has a generality, regardless of which
       chemical evolution model is used among different star formation
       rates with or without merging/outflow.
 \item By considering another source of iron supply by PISNe in Pop~III
       stars, we have deduced that the origin of $\mathrm{[Mg/Fe]}
       \sim -1$ in ULAS J1342 is a Pop~III PISN with $280~M_\sun$.  The
       mass of BLR required to reproduce $\mathrm{[Fe/H]} \sim +1$ is
       about $10^3~M_\sun$, which is consistent with the range suggested
       by the detailed photoionization simulation of
       \cite{2003ApJ...582..590B}.
 \item In the $\mathrm{[Mg/Fe]}\textendash z$ diagram, quasars of Pop
       III PISN origin are distributed vertically at high redshift,
       depending on the mass of the PISN progenitor, whereas quasars
       reflecting the chemical evolution of host elliptical galaxies are
       distributed horizontally.  These distributions are expected to
       intersect at $z = 6\textendash8$, and the previously considered
       [Mg/Fe] break may be buried in this intersection and difficult to
       identify.
 \item Another possible candidate for the iron supply to ULAS J1342 is
       the CCSN of a massive Pop~III star with $1000~M_\sun$.  In this
       case, however, the expected BLR mass would be $300~M_\sun$, which
       is much smaller than suggested by other studies.  In addition,
       cosmological simulations indicate that Pop~III stars with
       $>$1000~$M_\sun$ are extremely rare and their signature is
       unlikely to be found by chance in ULAS J1342.
 \item The silicon to iron abundance ratio [Si/Fe] in the BLR of ULAS
       J1342, if evaluated, enables {\color{black} one }to distinguish
       whether the heavy elements originated from a $280~M_\sun$ Pop~III
       PISN or a $1000~M_\sun$ Pop~III CCSN.  One candidate Si line to
       measure is \ion{Si}{4}~$\lambda1397$, which is the strongest
       silicon emission line seen in the UV spectrum of quasars.
 \item Various abundance ratios in the BLR are expected to show a Pop III PISN-characteristic, 
       such as a high ratio of [Si/Mg] for the large Si excess and 
       a low ratio of [Al/Mg] for the odd-even effect. All these heavy-element emission lines are 
       simultaneously observed in the near-infrared spectrum of QSOs at $z > 7$, 
       so that the [Si/Mg] and [Al/Mg] ratios for the BLR should provide additional
       confirmation of the existence of the Pop III PISNe. 
\end{enumerate}

The existence of a massive Pop~III star of PISN progenitor at high
redshift, if confirmed, may also impact ongoing
searches for signatures of massive Pop~III stars in the Galaxy. The VMP
stars with heavy-element abundance patterns of Pop~III PISN origin are
expected to have very low [Mg/Fe] compared to the observed average
for VMP stars in the halo. Discovering such stars by new generation
surveys may result in eventual derivation of the Pop~III IMF in the
Galaxy, complementing similar approaches based on future observational
studies of quasars at high redshifts.


\section*{acknowledgments}

We thank an anonymous referee for the careful and detailed review of our
 manuscript, which helped to clarify the presentation.  H.S. is
 supported by the Japan Society for the Promotion of Science (JSPS)
 KAKENHI Grant Number 19K03917 and 22K03683. T.T. is supported by JSPS
 KAKENHI 18H01258 and 19H05811, and T.S. by
 20H05639. T.C.B. acknowledges partial support from grant PHY 14-30152,
 Physics Frontier Center/JINA Center for the Evolution of the Elements
 (JINA-CEE), awarded by the US National Science Foundation, and the US
 National Science Foundation under Grant No. OISE-1927130 (IReNA).

Funding for the SDSS and SDSS-II has been provided by the Alfred
P. Sloan Foundation, the Participating Institutions, the National
Science Foundation, the U.S. Department of Energy, the National
Aeronautics and Space Administration, the Japanese Monbukagakusho, the
Max Planck Society, and the Higher Education Funding Council for
England. The SDSS Web Site is http://www.sdss.org/.

The SDSS is managed by the Astrophysical Research Consortium for the
Participating Institutions. The Participating Institutions are the
American Museum of Natural History, Astrophysical Institute Potsdam,
University of Basel, University of Cambridge, Case Western Reserve
University, University of Chicago, Drexel University, Fermilab, the
Institute for Advanced Study, the Japan Participation Group, Johns
Hopkins University, the Joint Institute for Nuclear Astrophysics, the
Kavli Institute for Particle Astrophysics and Cosmology, the Korean
Scientist Group, the Chinese Academy of Sciences (LAMOST), Los Alamos
National Laboratory, the Max-Planck-Institute for Astronomy (MPIA), the
Max-Planck-Institute for Astrophysics (MPA), New Mexico State
University, Ohio State University, University of Pittsburgh, University
of Portsmouth, Princeton University, the United States Naval
Observatory, and the University of Washington.


\bibliographystyle{apj}



\end{document}